\documentclass[twocolumn,preprintnumbers,amsmath,amssymb,a4paper]{revtex4-2}
\usepackage{bm}
\usepackage{extarrows}
\usepackage{amsfonts}
\usepackage{amsmath}
\usepackage{amssymb}
\usepackage{geometry}
\usepackage{graphicx}
\usepackage{footmisc}
\usepackage{braket}
\usepackage{array}
\usepackage{lipsum}
\usepackage{cancel}
\usepackage{bibentry}
\usepackage{natbib}
\usepackage[table,xcdraw]{xcolor}

\usepackage[colorlinks,linkcolor=blue,anchorcolor=blue,citecolor=blue,urlcolor=blue]{hyperref}

\begin{document}
	\title{Theoretical calculations of isotope shifts in highly charged Ni$^{12+}$ ion}

	\author{Shi-Cheng Yu$^{1,2}$}\email{yushicheng21@ucas.ac.cn}
	\author{Hua Guan$^{1,3,4}$}\email{guanhua@apm.ac.cn}
	\author{Lei She$^{1,3}$} \email{shelei@apm.ac.cn}
	\author{Cheng-Bin Li$^{1}$}\email{cbli@apm.ac.cn}
	\affiliation{$^1$Key Laboratory of Atomic Frequency Standards, Innovation Academy for Precision Measurement Science and Technology, Chinese Academy of Sciences, Wuhan 430071, China \\
		$^2$University of Chinese Academy of Sciences, Beijing 100049, China \\
		$^3$Wuhan Institute of Quantum Technology, Wuhan 430206, China \\
		$^4$Hefei National Laboratory, University of Science and Technology of China, Hefei, 230088, China
	}
	
	
	\begin{abstract}
		We present relativistic many-body perturbation theory plus configuration interaction (MBPT+CI) calculations of the lowest four excited states of Ni$^{12+}$, a promising candidate for highly charged ion (HCI) optical clocks. By combining the convergence behavior from multiple calculation models, we perform a detailed analysis of the electron-correlation effects and both the excitation energies and their uncertainties are obtained. Our computed energies for the first two excited states deviate from experimental values by less than $10~\mathrm{cm^{-1}}$, with relative uncertainties estimated below $0.2\%$. Building on the same computational procedure, we calculate the mass shift and field shift constants for the lowest four excited states of Ni$^{12+}$, and the resulting isotope shifts exhibit valence-correlation-induced relative uncertainties below the $1\%$ level.  These results provide essential atomic-structure input for high-precision isotope shift spectroscopy in Ni$^{12+}$.
	\end{abstract}
	
	\maketitle
	
	\section{Introduction}
	
	Optical atomic clocks have reached fractional frequency uncertainties at the $10^{-18}$--$10^{-19}$ level \cite{zhang_liquid-nitrogen-cooled_2025, al-2025, Yb-E3-2019, zhiqiang176LuClock2023, PhysRevLett.133.023401, JiaImprovedSystematicEvaluation2025, czlf-bfvp}, enabling a range of precision measurements that test fundamental physics \cite{filzingerImprovedLimitsCoupling2023, Yb-E3-2019, sr-mm-2022}. One notable application is isotope shift spectroscopy, where clock transitions in different isotopes of the same element are measured with extreme precision \cite{PhysRevLett.123.203001, PhysRevLett.125.123002, PhysRevLett.128.073001}. By constructing a King plot from such measurements, a linear relation between the isotope shifts of two transitions under the Standard Model of particle physics is expected, and a deviation from this linearity could signal the presence of a fifth force between electrons and neutrons, which may be mediated by a hypothetical light boson, provided that Standard Model sources of nonlinearity are well understood \cite{PhysRevA.110.L030801, PhysRevA.97.032510, PhysRevResearch.2.043444}.
	
	With the maturation of modern laser spectroscopy, isotope shift measurements are entering a regime where previously negligible contributions must be carefully controlled \cite{berengut2025precision}. This shifts the focus from purely experimental resolution toward identifying atomic systems that offer favorable intrinsic properties. In particular, optical clocks should provide narrow optical transitions while maintaining strong robustness against external-field-induced shifts. However, achieving such precision requires atomic systems with transitions that are both intrinsically narrow and minimally sensitive to external perturbations. 
    HCIs fulfill these requirements: the strong binding of their electrons results in compact orbitals, suppressing shifts from blackbody radiation, electric quadrupole interactions, and higher-order Zeeman effects \cite{kozlovHighlyChargedIons2018}. Furthermore, the nuclear–electron interaction is greatly enhanced in HCIs, increasing their sensitivity to variations in nuclear charge radii and the potential fifth force \cite{PhysRevLett.134.063002, PhysRevLett.134.233002}.
	
	Ni$^{12+}$ has attracted significant attention as an HCI clock candidate \cite{Yu-2018}. It possesses two visible-range forbidden transitions: a magnetic dipole ($M1$) transition at $\sim511$~nm and an electric quadrupole ($E2$) transition at $\sim498$~nm. The $E2$ transition, with a natural linewidth of only $8$~mHz \cite{Yu-2018}, is an excellent candidate for an optical clock transition, while the $M1$ transition can be used for quantum logic state detection. Nickel has several stable isotopes, making it an attractive system for King plot studies. 
	
	Theoretical and experimental studies on Ni$^{12+}$ have advanced rapidly in recent years. In particular, large-scale all-order plus CI calculations have enabled an unprecedented level of accuracy for a complex multivalent atomic system, yielding theoretical predictions of the $E2$ clock-transition energy with uncertainties below $10~\mathrm{cm^{-1}}$ \cite{flwf-c2m1}. This remarkable precision substantially narrowed the experimental search window and directly facilitated the subsequent high-precision measurement of the $E2$ transition frequency, which was determined with an experimental uncertainty of $0.01~\mathrm{cm^{-1}}$ \cite{flwf-c2m1}. In addition, the wavelength of the $M1$ transition has also been measured experimentally with an uncertainty of $0.018~\mathrm{cm^{-1}}$ using the Shanghai–Wuhan electron beam ion trap \cite{ChenPrecisionMeasurementM12024}. These advances have paved the way for the realization of a Ni$^{12+}$ optical clock, thereby enabling high-precision isotope shift measurements. However, isotope shift calculations for Ni$^{12+}$ is essential for interpreting future high-precision King plot measurements, have not yet been carried out.
	
	In this work, we perform the calculations of the energy levels and isotope shifts for the lowest four excited states ($^3P_1$, $^3P_0$, $^1D_2$ and $^1S_0$) in Ni$^{12+}$ using the MBPT+CI method. In our multireference CI calculations, we employee the emu CI method, which reduces the size of the CI space by directly discarding a large number of negligible CI matrix elements \cite{emuCI}. To validate its applicability to Ni$^{12+}$, we compare the results obtained with and without the emu CI method in the CI space with small $n_{\max}$, demonstrating that the emu CI method provides reliable accuracy while significantly improving computational efficiency. We combined the results from several multireference CI calculations to obtain the final energy levels and their associated uncertainties. For the lowest four excited states, the averaged relative uncertainty of our calculated energies is $0.14\%$, and the averaged relative deviation from the experimental values is $0.08\%$. This agreement demonstrates the reliability of both our computational model and the uncertainty evaluation procedure. Using the same computational model, we further calculated the mass shift and field shift constants for the lowest four excited states of Ni$^{12+}$, and the valence-correlation-induced relative uncertainties below the 1\% level.

	\section{method}\label{method}
	We carry out our MBPT+CI calculations using the AMBiT package, which is based on Brillouin–Wigner perturbation theory \cite{kahlAmbitProgrammeHighprecision2019}. The atomic wave function $\Psi$ with given total angular momentum $J$, projection $M$, and parity $P$ is expanded as a linear combination of configuration state functions (CSFs):
    \begin{eqnarray}
        \Psi(\gamma PJM) = \sum_i C_i \, \psi(\gamma_i PJM),
    \end{eqnarray}
    where $\psi(\gamma_i PJM)$ denotes a CSF constructed as a linear combination of Slater determinants. The coefficients of the CSFs are determined by the constraints on the total angular momentum, its projection, and the parity. Each Slater determinant is formed from the single-particle orbitals of the valence electrons.

    The single-electron orbitals are generated in the self-consistent Dirac–Hartree–Fock (DHF) potential derived from the Dirac–Coulomb–Breit (DCB) Hamiltonian:
    \begin{eqnarray}
        \hat{H}_\mathrm{DCB} = \sum_i \hat{h}_\mathrm{D}(i) + \sum_{i< j}\left(\frac{1}{r_{ij}} + B_{ij}\right),
    \end{eqnarray}
    where $B_{ij}$ is the Breit interaction and $\hat{h}_\mathrm{D}$ is the single-electron Dirac Hamiltonian:
    \begin{eqnarray}
        \hat{h}_\mathrm{D}(i) = &c \boldsymbol{\alpha}_i \cdot \mathbf{p}_i + (\beta_i - 1)mc^2 \nonumber\\
        &+ V_\mathrm{nuc}(r_i) + V_\mathrm{rad}(r_i).
    \end{eqnarray}
    Here, $V_\mathrm{rad}(r_i)$ is the QED radiative potential that accounts for the one-loop QED correction to the electron–nucleus interaction, while the electron–electron QED interaction is neglected \cite{radpot-PhysRevA.72.052115}. The nuclear potential $V_\mathrm{nuc}(r_i)$ includes the finite-size effect of the nucleus, modeled by a two-parameter Fermi charge distribution. The root-mean-square (rms) nuclear charge radius was taken as $R_\mathrm{rms} = 3.776~\mathrm{fm}$ for $^{58}$Ni.

    The CI+MBPT equation can then be written as:
    \begin{eqnarray}
        \sum_j \left[ H_{ij} + \Sigma^{(2)}_{ij}(E) \right] C_j = E C_i,
    \end{eqnarray}
    where $H_{ij}$ is the matrix element of the DCB Hamiltonian, and $\Sigma^{(2)}_{ij}(E)$ is the second-order self-energy correction representing virtual excitations of the core electrons.

    The ground configuration of Ni$^{12+}$ is [Ne]$3s^2 3p^4$. In our calculations, the $3s^2 3p^4$ electrons are treated as valence electrons, and the core orbitals and virtual orbitals are generated from the DHF calculation in $V^{N-6}$ potential (without valence electrons). The single-electron basis is constructed from 40 B-splines of order $k=10$ confined within a spherical cavity of radius $R_\mathrm{max} = 5\,a_0$. Orbitals with orbital angular momentum $\ell \leq 4$ and principal quantum number $n \leq 30$ are included in the evaluation of the $\Sigma^{(2)}_{ij}(E)$ operator.

	\begin{figure}[htbp]
	    \centering
	    \includegraphics[width=\linewidth]{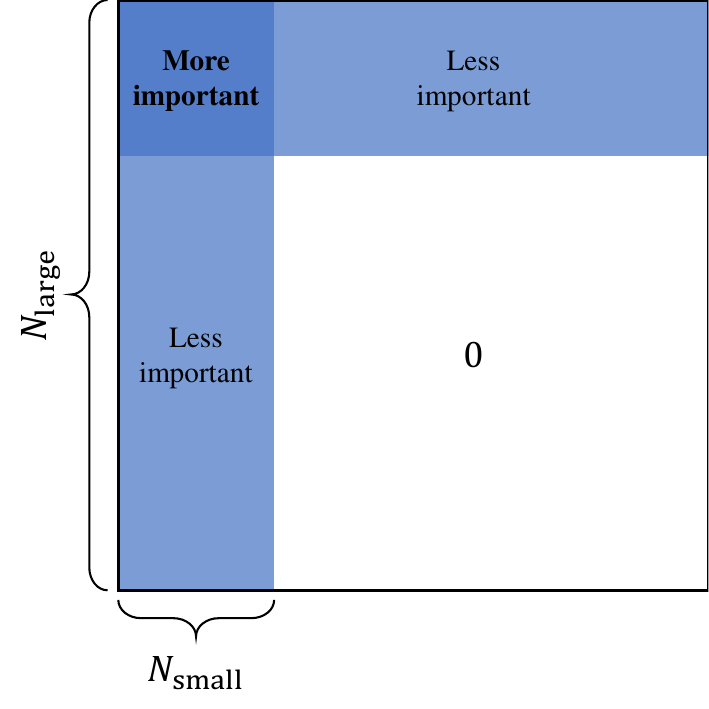}
	    \caption{Schematic diagram of the CI matrix in the emu CI method. $N_\mathrm{small}$ denotes the size of the small CI subspace containing the more important configurations, while $N_\mathrm{large}$ represents the dimension of the original CI space.}
	    \label{emu}
	\end{figure}
	
	The CI space is constructed by including all single and double excitations from the chosen reference configurations, with truncations applied to the principal quantum number $n_{\max}$ and partial-wave $\ell_{\max}$. Increasing these cutoffs systematically improves the accuracy of the calculation and allows the results to approach the complete basis set (CBS) limit. However, this also leads to a rapid increase in the dimension of the CI space, making the calculations computationally demanding. To address this problem, we employed the emu CI method to reduce the size of the CI space. As illustrated in Fig.~\ref{emu}, the emu CI method defines a small CI subspace that contains the important configurations. Only the CI matrix elements that couple to this small subspace are retained, while all other matrix elements are set to zero. This approach substantially reduces the computational cost while maintaining high accuracy, and has been successfully applied in various MBPT+CI calculations \cite{yuPotentialIons22024, LiuSpectralPropertiesPolarizabilities2021, BekkerDetection5p4fOrbital2019, PhysRevA.109.023106}.

	\section{Results}\label{res}

    In this section, we present the results of our calculations and the corresponding uncertainty analysis. We first outline the procedure adopted to estimate the theoretical uncertainties in the MBPT+CI calculations in the first subsection. The following two subsections report the calculated energy levels of Ni$^{12+}$ and the isotope shifts among its stable isotopes, respectively. 

	\subsection{Uncertainty estimations}
	Owing to the high ionization degree of the Ni$^{12+}$ ion, the core–valence correlations are relatively small. The difference between the all-order method and the MBPT method is less than $10~\mathrm{cm^{-1}}$, as reported in Ref.~\cite{flwf-c2m1}. Therefore, we only consider the uncertainties arising from the CI calculations.

    In detailed, we treat the three-reference emu CI model as the basis model, and three corrections are taken into account:

    (1) Basis-set convergence.  
    We enlarged the truncation limits of the principal quantum number $n_{\max}$ and partial-wave $\ell_{\max}$ in the single-reference  CI calculations (where the non-relativistic reference configuration is $3s^2 3p^4$) to approach the CBS limit. The correction and corresponding uncertainty are denoted as $\Delta_{\mathrm{CBS}}$ and $\sigma_{\mathrm{CBS}}$, respectively.

    (2) Number of reference configurations. 
    To assess the effect of the reference configuration selection, we compared the results of emu CI calculations using three-reference (where the non-relativistic reference configurations are $3s^2 3p^4, 3s^2 3p^2 3d^2, 3s^1 3p^4 3d^1$) and five-reference (where the non-relativistic reference configurations are $3s^2 3p^4, 3s^2 3p^2 3d^2, 3s^1 3p^4 3d^1, 3s^2 3p^3 4p^1, 3s^1 3p^4 4s^1$). The correction and corresponding uncertainty are denoted as $\Delta_{\mathrm{ref}}$ and $\sigma_{\mathrm{ref}}$, respectively.

    (3) The emu CI approximation.  
    In the three-reference CI framework, we performed calculations with and without employing the emu CI method, where the small CI subspace is constructed by including all single and double excitations from the reference configuration $3s^2 3p^4$. The correction and corresponding uncertainty are denoted as $\Delta_{\mathrm{emu}}$ and $\sigma_{\mathrm{emu}}$, respectively.

    These differences are treated as corrections to the results obtained from the three-reference emu CI calculations, and are used to estimate the theoretical uncertainty $\sigma$:
    \begin{eqnarray}
        \sigma = \sqrt{\sigma_{\mathrm{CBS}}^2 + \sigma_{\mathrm{ref}}^2 + \sigma_{\mathrm{emu}}^2}.
    \end{eqnarray}

	\subsection{Energy levels}
	
	\begin{figure*}[htbp]
	    \centering
	    \includegraphics[width=\linewidth]{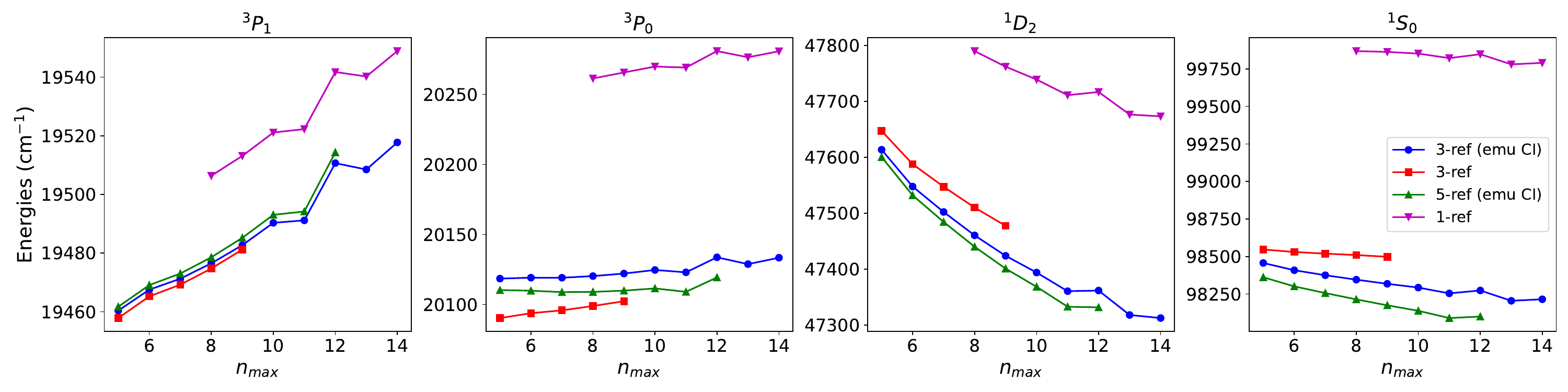}
	    \caption{Convergence of the calculated energies of the lowest four excited states in different computational models as a function of the principal quantum number cutoff $n_{\max}$, with the partial-wave cutoff fixed at $\ell_{\max} = 4$.}
	    \label{fig:energy_convergence}
	\end{figure*}
	\begin{figure*}[htbp]
	    \centering
	    \includegraphics[width=\linewidth]{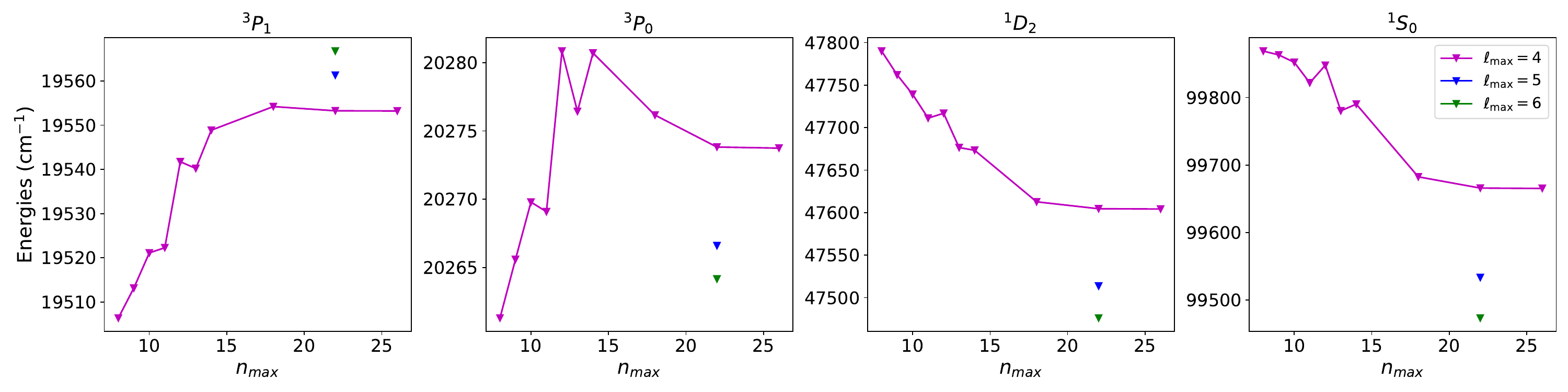}
	    \caption{Convergence of the calculated energies of the lowest four excited states in single-reference CI results as a function of the principal quantum number cutoff $n_{\max}$ and partial-wave cutoff $\ell_{\max}$.}
	    \label{fig:sr}
	\end{figure*}
	
	To examine the convergence behavior and reliability of our CI calculations, we compare the energies of the lowest four excited states obtained from multiple computational models. In particular, we investigate the dependence of the calculated energies on the principal quantum number cutoff $n_{\max}$ while keeping the partial-wave cutoff fixed at $\ell_{\max}=4$. The multiple computational models correspond to varying the number of reference configurations and the use of the emu CI method. 
	
	
	As shown in Fig.~\ref{fig:energy_convergence}, apart from the single-reference CI calculations, the results obtained from the three-reference emu CI and five-reference emu CI models are in close agreement, indicating that the CI space constructed from the three reference configurations already includes the dominant configurations contributing to the energies of the lowest four excited states. The inclusion of additional reference configurations introduces only minor changes. We also observe that the deviations between the two models tend to increase slightly with $n_{\max}$. For the lowest four excited states, the maximum relative deviations occur at $n_{\max}=12$, being 0.02\%, 0.07\%, 0.06\%, and 0.18\%, respectively, which demonstrates the adequacy of the three-reference configuration choice.

    In addition, we examined the effect of the emu CI method at small principal quantum number cutoff $n_{\max}$. For the $^3\!P_1$ and $^3\!P_0$ states, the variations introduced by the emu CI method decrease with increasing $n_{\max}$, reaching relative deviations of 0.01\% and 0.10\% at $n_{\max}=9$. In contrast, for the $^1\!D_2$ and $^1\!S_0$ states, the deviations increase slightly with $n_{\max}$, with relative variations of 0.11\% and 0.18\% at $n_{\max}=9$. In the three-reference CI calculations, the emu CI method retains less than 3\% of the total CI matrix elements compared with the normal CI calculation, leading to a substantial reduction in computational time and memory requirements. These comparisons confirm the effectiveness of the emu CI method in providing accurate results while significantly reducing the computational cost. 

    Therefore, since the emu CI method proves to be sufficiently accurate and the inclusion of additional reference configurations does not produce significant contributions, we adopt the three-reference emu CI calculations with $n_{\max}=14$ and $\ell_{\max}=4$ as our basis model. The correction  $\Delta_{\mathrm{ref}}$, is defined as the difference between the results of the five-reference and three-reference emu CI calculations at $n_{\max}=12$ and $\ell_{\max}=4$, and the uncertainty $\sigma_{\mathrm{ref}} = \Delta_{\mathrm{ref}}$. The correction $\Delta_{\mathrm{emu}}$, is defined as the difference between the results of the three-reference CI and three-reference emu CI calculations at $n_{\max}=9$ and $\ell_{\max}=4$, and the uncertainty $\sigma_{\mathrm{emu}} = \Delta_{\mathrm{emu}}$.

    Although the single-reference CI results deviate noticeably from those obtained with multi-reference models, the deviations remain relatively stable and show no clear trend with increasing $n_{\max}$, as illustrated in Fig.~\ref{fig:energy_convergence}. Moreover, the single-reference CI calculations require substantially less memory, allowing us to extend both $n_{\max}$ and $\ell_{\max}$ to larger values. This enables us to approach the CBS limit and to analyze the contribution from basis-set convergence.

    The convergence of the single-reference CI results with respect to $n_{\max}$ and $\ell_{\max}$ is shown in Fig.~\ref{fig:sr}. We observe that at $\ell_{\max}=4$, the energies are already converged when the principal quantum number cutoff reaches $n_{\max}=22$, and further increasing $n_{\max}$ changes the energies of the lowest four excited states by less than $0.6~\mathrm{cm^{-1}}$. This behavior differs from that reported in Ref.~\cite{flwf-c2m1}, which may originate from differences in the nodal structure of the basis functions and the choice of cavity radius $R_{\max}$. The CBS correction, $\Delta_{\mathrm{CBS}}$, is defined as the difference between the single-reference CI results obtained with $n_{\max}=14$, $\ell_{\max}=4$ and those obtained with $n_{\max}=22$, $\ell_{\max}=6$. To quantify the associated uncertainty, we decompose the CBS uncertainty $\sigma_{\mathrm{CBS}}$ into independent contributions from $n_{\max}$ and $\ell_{\max}$. The uncertainty related to $n_{\max}$, denoted as $\sigma_{\mathrm{CBS}}(n_{\max})$, is taken as the difference between the single-reference CI results obtained with $n_{\max}=22$, $\ell_{\max}=4$ and with $n_{\max}=18$, $\ell_{\max}=4$. The uncertainty associated with $\ell_{\max}$, denoted as $\sigma_{\mathrm{CBS}}(\ell_{\max})$, is defined as the difference between the results with $n_{\max}=22$, $\ell_{\max}=6$ and with $n_{\max}=22$, $\ell_{\max}=5$. The total CBS uncertainty is then given by the quadrature sum:
    \begin{eqnarray}
    \sigma_{\mathrm{CBS}} = \sqrt{ \sigma_{\mathrm{CBS}}(n_{\max})^{2} + \sigma_{\mathrm{CBS}}(\ell_{\max})^{2} }.
    \end{eqnarray}

    Compared with typical MBPT+CI calculations for HCIs with fewer valence electrons, a significantly larger $n_{\max}$ and $\ell_{\max}$ are required here to achieve convergence. This behavior arises because the DHF orbitals generated in the $V^{N-6}$ potential, where the valence electrons are excluded from the self-consistent field, are not sufficiently close to the true valence orbitals. When the MBPT+CI calculation is performed using orbitals obtained in the $V^{N-6}$ potential, the weight of the $3s^2 3p^4$ configuration in the wave function is about 90\%. In contrast, if the DHF orbitals are generated in the $V^{N}$ potential, which includes the valence electrons, the corresponding configuration weight exceeds 96\%, and the convergence with respect to $n_{\max}$ and $\ell_{\max}$ is considerably faster. This comparison supports our interpretation regarding the influence of the DHF potential on the basis quality.

    However, within the MBPT+CI framework, the use of the $V^{N}$ potential requires the inclusion of additional subtraction diagrams and leads to an enhanced contribution from core–valence correlations. In the absence of a comparison between all-order and MBPT calculations performed in the $V^{N}$ potential, the uncertainty associated with the treatment of core–valence correlations is difficult to assess. For this reason, we perform our calculations in the $V^{N-6}$ potential, ensuring that the core–valence correlations are treated as accurately as possible, at the cost of increased CI-space complexity and slower convergence.

    \begin{table}[htbp]
        \centering
        \begin{ruledtabular}
        \begin{tabular}{lrrrr}
                                & \multicolumn{1}{r}{$^3P_1$}   & \multicolumn{1}{r}{$^3P_0$} & \multicolumn{1}{r}{$^1D_2$} & \multicolumn{1}{r}{$^1S_0$} \\ \hline
        Basis                   & 19517.74                      & 20133.32                    & 47312.52                    & 98214.82                    \\
        $\Delta_{\mathrm{CBS}}$ & 17.90                         & $-$16.54                      & $-$197.33                     & $-$316.75                     \\
        $\Delta_{\mathrm{ref}}$ & 3.68                          & $-$14.38                      & $-$30.10                      & $-$173.68                     \\
        $\Delta_{\mathrm{emu}}$ & $-$1.49                         & $-$19.75                      & 53.66                       & 180.05                      \\
        MBPT$^*$                    & 193.91                       &  42.82                    &   $-$176.68                   &  $-$582.67                   \\
        QED$^*$                 &  52.11                     &   33.68            &   44.43              &    52.65                \\
        Final                   & 19538(7)                     & 20083(25)                   & 47139(73)                  & 97904(258)                  \\ 
        & & & & \\
        Year             & \multicolumn{4}{c}{Other theoretical results} \\\hline
        2018\footnote{MCDHF calculation \cite{Yu-2018}}                     & 19560(20)                     & 20251(450)                  & 47512(2120)                 & 99377(5130)                 \\
        2018\footnote{MCDHF calculation \cite{Wang_2018}} & 19534 & 20140 & 47189 & 98121 \\
        2021\footnote{MCDHF calculation \cite{Liang-2021}} & 19540(21)  & 20125(97) \\
        2024\footnote{MCDHF calculation \cite{Chen-2024}}                     & \multicolumn{1}{l}{19551(10)} & \multicolumn{1}{l}{}        & \multicolumn{1}{l}{}        & \multicolumn{1}{l}{}        \\
        2025\footnote{all-order and pure CI calculations \cite{flwf-c2m1}}                    & 19547                         & 20086                       & 47063                       & 97894                       \\
                                & 19550                         & 20081                       & 47051                       & 97771                       \\ \hline
        Expt.                    & 19542\footnote{\cite{Chen-2024}}                         & 20079\footnote{\cite{flwf-c2m1}}                       & 47033\footnote{\cite{NIST}}                       & 97836\footnote{\cite{NIST}}                       
        \end{tabular}
        \end{ruledtabular}
        \caption{Calculated energies (in cm$^{-1}$) of the lowest four excited states of Ni$^{12+}$. Listed are the basis results, corrections from $\Delta_{\mathrm{CBS}}$, $\Delta_{\mathrm{ref}}$, and $\Delta_{\mathrm{emu}}$, and the MBPT and QED contributions. The final values are compared with experimental results (Expt.) and other theoretical results. An asterisk (*) in the row label indicates that the corresponding effect is already included in the basis calculation, and the listed values are the difference between calculations with and without that effect. Numbers in parentheses denote the uncertainties.}
        \label{tab:final-el}
    \end{table}

	Table~\ref{tab:final-el} summarizes the final calculated energies after including all corrections. The basis model corresponds to the three-reference emu CI calculations with $n_{\max}=14$ and $\ell_{\max}=4$, while the definitions and estimations of the individual corrections and their associated uncertainties are discussed in the previous discussion. Among the three corrections, $\Delta_{\mathrm{CBS}}$ is found to be the most significant, providing the largest contribution for all states except the $^3\!P_0$ level. This confirms the reliability of our chosen basis model, which includes the most important three reference configurations to ensure that the CI space contains all dominant configurations. The use of the emu CI method effectively reduces the computational cost without compromising accuracy, allowing us to extend $n_{\max}$ and thereby approach the CBS limit. For the $^3\!P_1$ level, the dominant source of uncertainty is $\sigma(\ell_{\max})$, whereas for the $^3\!P_0$, $^1\!D_2$, and $^1\!S_0$ levels, the uncertainties are primarily determined by $\sigma_{\mathrm{emu}}$ and $\sigma_{\mathrm{ref}}$, with the emu CI contribution being the more significant of the two.

    We also compare our results with several theoretical and experimental results. The mean relative deviation from the experimental energies is only 0.08\%, and the estimated mean relative uncertainty is 0.14\%, which is larger than these deviations, indicating that our uncertainty evaluation is conservative and reliable. For the $^1\!D_2$ level, the deviation between our theoretical result and the experimental result corresponds to approximately $1.5\sigma$. This discrepancy may originate from an insufficiently accurate estimation of the dominant contribution $\Delta_{\mathrm{emu}}$, which exhibits a tendency to increase as $n_{\max}$ is enlarged. However, further extension of $n_{\max}$ is currently limited by the available computational memory. Despite this, the overall agreement between our theoretical results and experimental results remains very good.
    Overall, our results show both higher accuracy and smaller uncertainties than those obtained from previous Multiconfiguration Dirac–Hartree–Fock (MCDHF) calculations. 

	High-precision all-order+CI and 16-electron pure CI calculations were performed in Ref.~\cite{flwf-c2m1}. Our MBPT+CI results show excellent agreement with their all-order+CI values for three of the excited states, with differences not exceeding $10~\mathrm{cm^{-1}}$, indicating that the higher-order core–valence correlations are sufficiently small for these levels. In the calculation of QED corrections, our results are also in very good agreement with those reported. In Ref.~\cite{flwf-c2m1}, the nonlocal model QED operator (implemented as the QEDMOD package in Refs.~\cite{SHABAEV2018, PhysRevA.88.012513}) was incorporated into the all-order+CI framework, yielding QED shifts of 50, 33, 43, and 50~cm$^{-1}$ for the lowest four excited states. These values differ from our results obtained using the QED radiative potential by only a few percent, demonstrating the good consistency between the two approaches to evaluating QED effects in HCIs within middle-Z region. The small remaining differences may arise from the treatment of electron correlations, as Ref.~\cite{flwf-c2m1} also included the QEDMOD operator in the 16-electron pure CI calculation, which gave corresponding QED shifts of 49, 33, 42, and 52~cm$^{-1}$ for the lowest four excited states, slightly different from those obtained in their all-order+CI calculation. Given the consistent agreement among the three independent evaluations of the QED shifts, we conclude that the uncertainty associated with the QED potential in the energy level calculations is negligible.

	\subsection{Isotope shifts}

    The isotope shift of energy levels arises from two main contributions: 
    the finite nuclear mass (mass shift) and the finite size of the nuclear charge distribution (field shift). 
    Thus, the total isotope shift between two isotopes with mass numbers $A$ and $A'$ can be expressed as
    \begin{eqnarray}
    \frac{\delta E^{A',A}}{h} = k\left(\frac{1}{A'} - \frac{1}{A}\right) + F_a\,\delta\langle r^2\rangle^{A',A},
    \end{eqnarray}
    where $k$ is the mass shift constant associated with the change in the nuclear mass, and 
    $F_a$ is the field shift constant corresponding to the variation of the mean-square nuclear charge radius $\delta\langle r^2\rangle^{A',A}$.
        
    The mass shift can be divided into two parts: the normal mass shift (NMS) and the specific mass shift (SMS):
    \begin{equation}
    k = k_{\mathrm{NMS}} + k_{\mathrm{SMS}},
    \end{equation}
    both contributions can be evaluated self-consistently within a finite-field approach by introducing a scaling parameter $\lambda$ into the atomic Hamiltonian,
    \begin{equation}
    \hat{H}(\lambda) = \hat{H}_0 + \lambda\,\hat{H}_{\mathrm{NMS/SMS}},
    \end{equation}
    where $\hat{H}_0$ is the MBPT+CI Hamiltonian. The relativistic NMS and SMS operators are
    \begin{eqnarray}
    \hat{H}_\mathrm{NMS} &=& 
    \sum_{i} \Big[ \mathbf{p}_i^2 
    - \frac{\alpha Z}{r_i} \boldsymbol{\alpha}_i \cdot \mathbf{p}_i  \nonumber\\
    && \quad - \frac{\alpha Z}{r_i} (\boldsymbol{\alpha}_i \cdot \mathbf{C}^1_i)\,
    \mathbf{C}^1_i \cdot \mathbf{p}_i \Big], \\
    \hat{H}_\mathrm{SMS} &=& 
    \sum_{i\neq j} \Big[ \mathbf{p}_i \cdot \mathbf{p}_j 
    - \frac{\alpha Z}{r_i} \boldsymbol{\alpha}_i \cdot \mathbf{p}_j \nonumber\\
    && \quad - \frac{\alpha Z}{r_i} (\boldsymbol{\alpha}_i \cdot \mathbf{C}^1_i)\,
    \mathbf{C}^1_i \cdot \mathbf{p}_j \Big],
    \end{eqnarray}
    and the constant $k_{\mathrm{NMS,SMS}}$ is obtained from the finite-field derivative:
    \begin{equation}
    k_{\mathrm{NMS/SMS}} = \left.\frac{dE(\lambda)}{d\lambda}\right|_{\lambda=0}.
    \end{equation}
    
    The field shift constant $F_a$ can be evaluated analogously by varying the nuclear charge radius $\langle r^2 \rangle$ in the nuclear potential $V_{\mathrm{nuc}}$ and computing
    \begin{equation}
    F_a = \frac{\langle \delta V_{\mathrm{nuc}} \rangle}{\delta\langle r^2\rangle}. 
    \end{equation}
    
    The random phase approximation (RPA) is used in the calculation of matrix element $\langle \delta V_{\mathrm{nuc}} \rangle$.
    
    \begin{figure*}[!htbp]
	    \centering
	    \includegraphics[width=\linewidth]{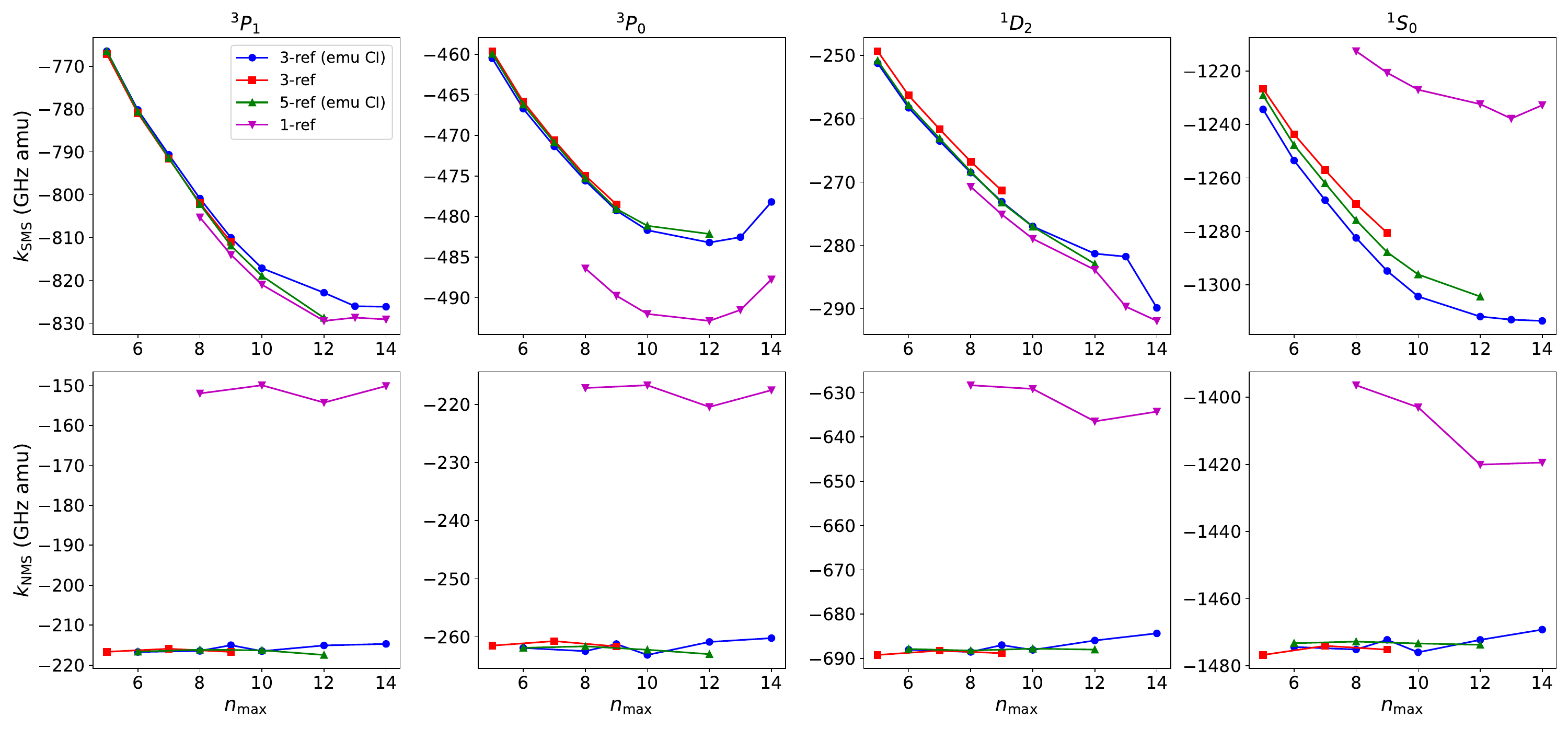}
	    \caption{Convergence of the calculated $k_{\mathrm{SMS}}$ and $k_{\mathrm{NMS}}$ of the lowest four excited states in different computational models as a function of the principal quantum number cutoff $n_{\max}$, with the partial-wave cutoff fixed at $\ell_{\max} = 4$.}
	    \label{fig:k_convergence}
	\end{figure*}
	
	\begin{figure*}[!htbp]
	    \centering
	    \includegraphics[width=\linewidth]{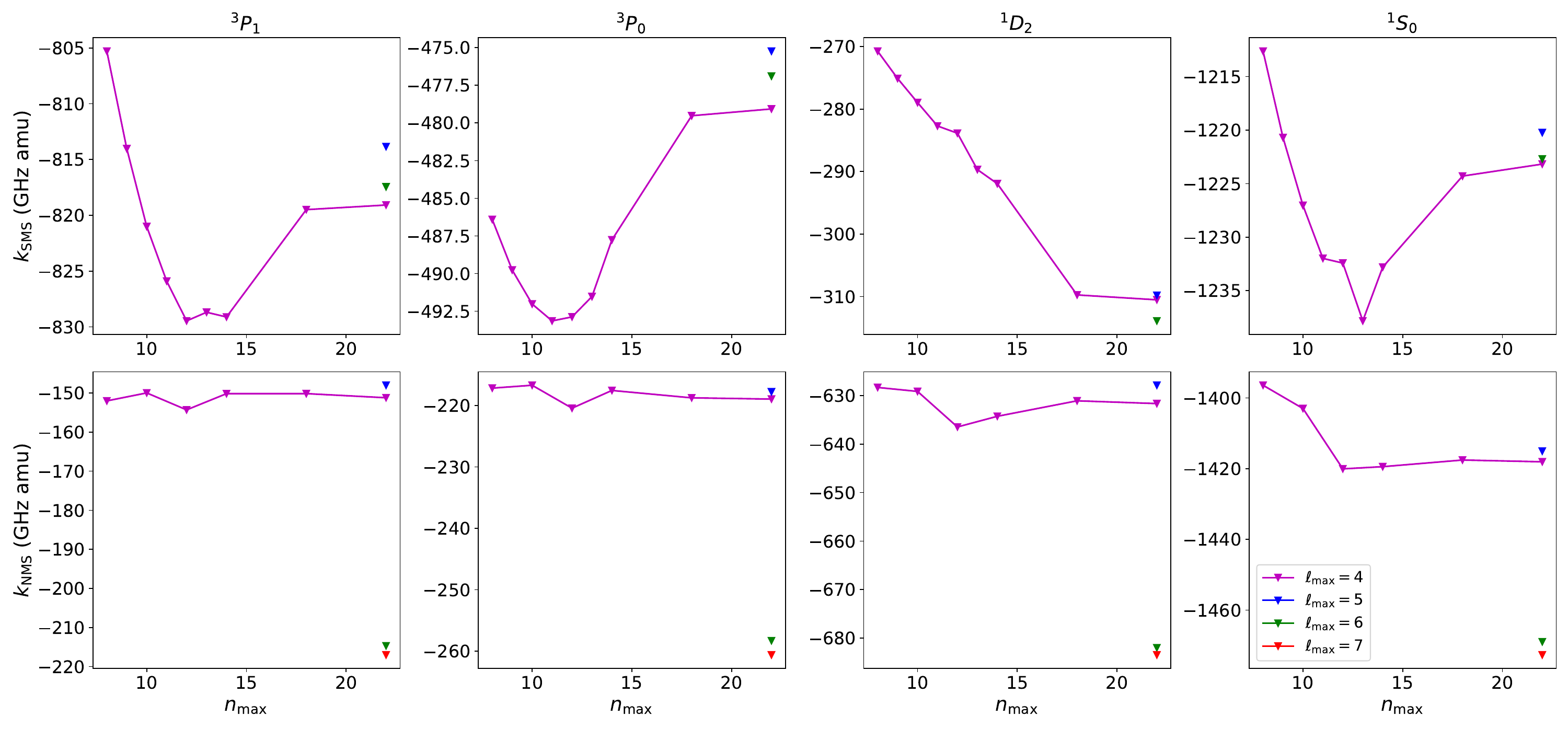}
	    \caption{Convergence of the calculated $k_{\mathrm{SMS}}$ and $k_{\mathrm{NMS}}$ of the lowest four excited states in single-reference CI results as a function of the principal quantum number cutoff $n_{\max}$ and partial-wave cutoff $\ell_{\max}$.}
	    \label{fig:k_1ref}
	\end{figure*}
    
    Based on the theoretical framework described above, we calculate the mass shift constants $k_{\mathrm{SMS}}$ and $k_{\mathrm{NMS}}$ as well as the field shift constant $F_a$, following the same computational procedure used for the energy levels. We first discuss the mass shift constants. The convergence of $k_{\mathrm{SMS}}$ and $k_{\mathrm{NMS}}$ under multiple calculation models is shown in Fig.~\ref{fig:k_convergence}, while the convergence behavior under the single-reference CI model is presented in Fig.~\ref{fig:k_1ref}. For $k_{\mathrm{SMS}}$, the convergence curves in Fig.~\ref{fig:k_convergence} exhibit features similar to those observed for the energy levels: the differences between the three-reference CI and the single-reference CI calculations remain relatively stable and show similar trends across $n_{\max}$. In addition, the three models excluding the single-reference CI produce results that are in very good agreement with each other, indicating that $k_{\mathrm{SMS}}$ is not highly sensitive to the variation of emu CI method or to the inclusion of additional reference configurations. Figure~\ref{fig:k_1ref} shows the behavior of $k_{\mathrm{SMS}}$ at large $n_{\max}$ and $\ell_{\max}$. The results converge at $n_{\max}=22$, and the extension of $\ell_{\max}$ has only a minor impact on the final values.

    In contrast to $k_{\mathrm{SMS}}$, the behavior of $k_{\mathrm{NMS}}$ is markedly different. As shown in Fig.~\ref{fig:k_convergence}, all calculation models exhibit only a weak dependence on $n_{\max}$, while the differences between the three-reference CI and the single-reference CI calculations are significantly larger than those observed for $k_{\mathrm{SMS}}$. Nevertheless, the three models excluding the single-reference CI still remain in good mutual agreement. Figure~\ref{fig:k_1ref} further illustrates that the convergence with respect to $n_{\max}$ is considerably faster for $k_{\mathrm{NMS}}$ than for $k_{\mathrm{SMS}}$, whereas the extension of $\ell_{\max}$ introduces a substantial correction to the final values. Specifically, the correction from $\ell_{\max}=4$ to $\ell_{\max}=5$ is very small, while the change from $\ell_{\max}=5$ to $\ell_{\max}=6$ is significant, amounting to approximately $-50$~GHz~amu. To more thoroughly investigate the convergence behavior at high $\ell_{\max}$, we extend the single-reference CI calculations of $k_{\mathrm{NMS}}$ to $\ell_{\max}=7$. We find that the results are well converged at this level, with the extension from $\ell_{\max}=6$ to $\ell_{\max}=7$ altering $k_{\mathrm{NMS}}$ by only about 2~GHz~amu. In this case, the uncertainty $\sigma_{\mathrm{CBS}}(\ell_{\max})$ is defined as the difference between the results obtained with $\ell_{\max}=7$ and $\ell_{\max}=6$ at $n_{\max}=22$.
    
    Table~\ref{tab:final-k} summarizes the final calculated mass shift constants after including all corrections. For the $^3\!P_1$, $^3\!P_0$, and $^1\!D_2$ states, the uncertainties in $k_{\mathrm{SMS}}$ are dominated by $\sigma(\ell_{\max})$, whereas for the $^1\!S_0$ state, the dominant contribution arises from $\sigma_{\mathrm{emu}}$. The $k_{\mathrm{SMS}}$ results obtained from the basis model are already close to the final results that include all corrections, with differences at the level of only a few percent. Nevertheless, to achieve uncertainties below the 1\% level, the inclusion of the correction terms is essential, particularly the evaluation of $\Delta_{\mathrm{CBS}}$ and $\sigma_{\mathrm{CBS}}$. 

    For $k_{\mathrm{NMS}}$, the CBS correction plays an even more prominent role: the relative correction to the basis model results can exceed 10\%, and this contribution originates almost entirely from the inclusion of the $\ell=6$ partial-wave. This highlights the importance of combining the convergence behavior obtained from multiple calculation models in order to produce reliable final values. For systems such as Ni$^{12+}$, which involve a large number of valence electrons, only calculation models with a small number of reference configurations can feasibly reach partial-wave cutoff as high as $\ell_{\max}=6$, making this multi-model convergence strategy particularly crucial. Additionally, in contrast to $k_{\mathrm{SMS}}$, the $k_{\mathrm{NMS}}$ results exhibit weak sensitivity to the MBPT corrections.

    \begin{table}[!htbp]
        \centering
        \begin{ruledtabular}
        \begin{tabular}{lrrrr}
         $k_{\mathrm{SMS}}$    & \multicolumn{1}{r}{$^3P_1$}   & \multicolumn{1}{r}{$^3P_0$} & \multicolumn{1}{r}{$^1D_2$} & \multicolumn{1}{r}{$^1S_0$} \\ \hline
        Basis                   & $-826.16$          & $-478.21$    & $-289.88$    & $-1313.47$                 \\
        $\Delta_{\mathrm{CBS}}$ & $11.67$            & $10.87$      & $-21.94$     & 10.10                  \\
        $\Delta_{\mathrm{ref}}$ & $-2.73$            & $0.40$       & $-1.15$      & 8.64                    \\
        $\Delta_{\mathrm{emu}}$ & $-1.00$            & 0.77         & 1.75         & 14.25                    \\
        MBPT$^*$                & $-46.46$           & $3.68$       & $16.69$      & 124.92            \\
        QED$^*$                 & $5.03  $           & $4.26 $      & $ 3.87$      & 1.73                  \\
        Final                   & $-818(5)$          & $-466(2)$    & $-311(5)$    & $-1280(17)$                 \\ 
        & & & & \\
        $k_{\mathrm{NMS}}$         & \multicolumn{1}{r}{$^3P_1$}   & \multicolumn{1}{r}{$^3P_0$} & \multicolumn{1}{r}{$^1D_2$} & \multicolumn{1}{r}{$^1S_0$} \\ \hline
        Basis                   & $-214.70$          & $-260.24$    & $-684.33$    & $-1469.21$                 \\
        $\Delta_{\mathrm{CBS}}$ & $-64.52$           & $-40.76$     & $-47.71$     & $-49.52$                  \\
        $\Delta_{\mathrm{ref}}$ & $-2.37$            & $-2.10$      & $-2.06$      & $-1.46$                  \\
        $\Delta_{\mathrm{emu}}$ & $-1.67$            & $-0.41$      & $-1.82$      & $-2.92$                    \\
        MBPT$^*$                & $-1.71$            & $-1.27$      & 4.00         & 12.47            \\
        QED$^*$                 & $-0.48$            & $-1.01$      & $-0.70$      & $-1.40$                  \\
        Final                   & $-286(4)$          & $-306(3)$    & $-737(3)$    &  $-1527(5)$  
        \end{tabular}
        \end{ruledtabular}
        \caption{Calculated mass shift constants (in GHz amu) of the lowest four excited states of Ni$^{12+}$. Listed are the basis results, corrections from $\Delta_{\mathrm{CBS}}$, $\Delta_{\mathrm{ref}}$, and $\Delta_{\mathrm{emu}}$, and the MBPT and QED contributions. An asterisk (*) in the row label indicates that the corresponding effect is already included in the basis calculation, and the listed values are the difference between calculations with and without that effect. Numbers in parentheses denote the uncertainties.}
        \label{tab:final-k}
    \end{table}

\begin{figure*}[htbp]
	    \centering
	    \includegraphics[width=\linewidth]{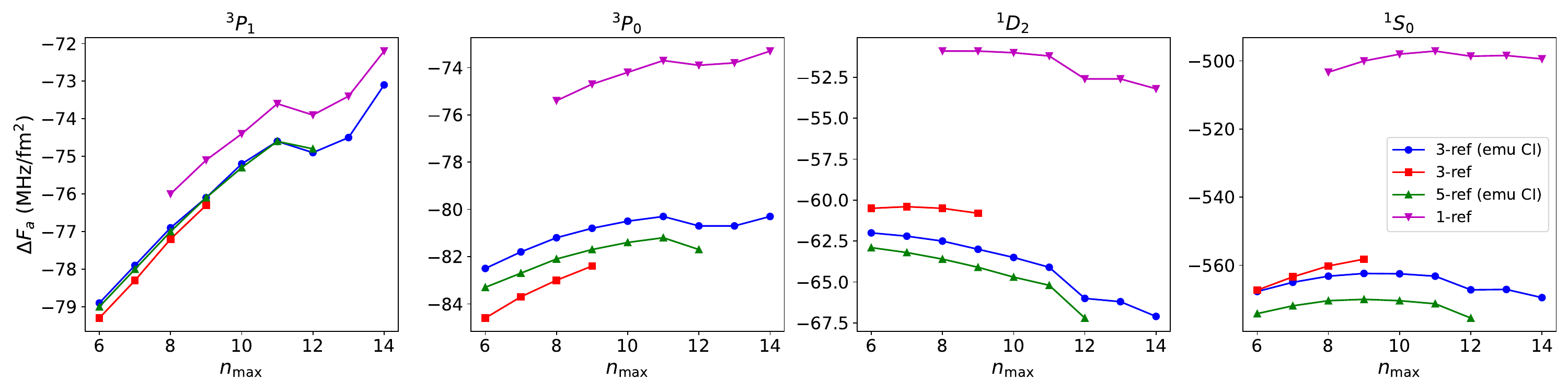}
	    \caption{Convergence of the calculated $\Delta F_a$ of the lowest four excited states in different computational models as a function of the principal quantum number cutoff $n_{\max}$, with the partial-wave cutoff fixed at $\ell_{\max} = 4$.}
	    \label{fig:fs_convergence}
	\end{figure*}

\begin{figure*}[htbp]
	    \centering
	    \includegraphics[width=\linewidth]{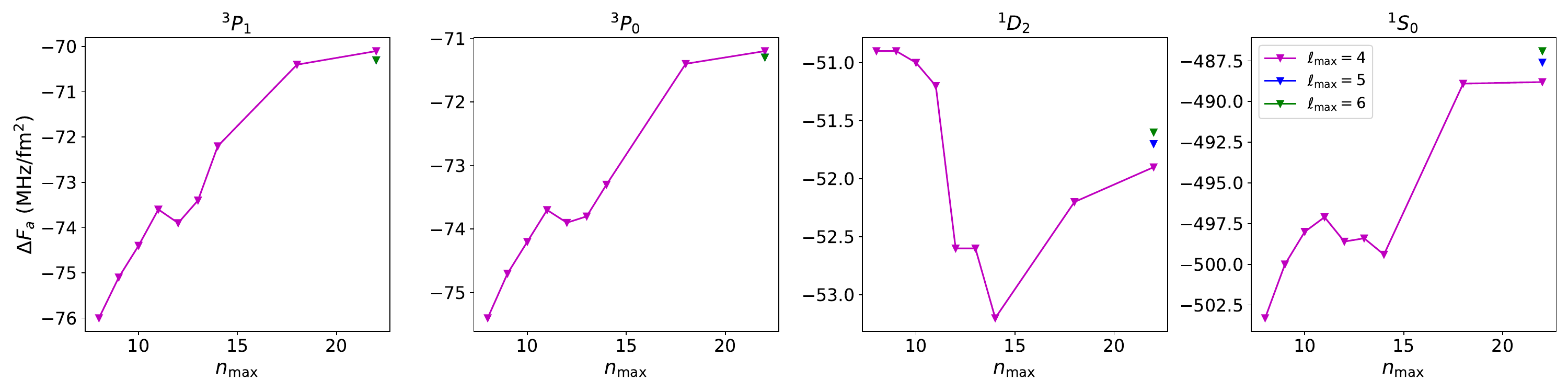}
	    \caption{Convergence of the calculated $\Delta F_a$ of the lowest four excited states in single-reference CI results as a function of the principal quantum number cutoff $n_{\max}$ and partial-wave cutoff $\ell_{\max}$.}
	    \label{fig:fs_1ref}
	\end{figure*}

	Since our primary interest lies in the measurable transition frequencies rather than the absolute energies, the discussion of field shift constants focuses on the differential field shift constants $\Delta F_a$, defined as the field shift constants with respect to the ground state $^3P_2$. The convergence of $\Delta F_a$ with respect to $n_{\max}$ under different calculation models is shown in Fig.~\ref{fig:fs_convergence}, while the convergence behavior under the single-reference CI model with varying $n_{\max}$ and $\ell_{\max}$ is presented in Fig.~\ref{fig:fs_1ref}. The convergence characteristics of $\Delta F_a$ are similar to those observed for the energy levels and for $k_{\mathrm{SMS}}$: the three-reference CI and single-reference CI calculations exhibit similar trends as $n_{\max}$ increases, and the single-reference CI results reach convergence around $n_{\max}=22$.

	\begin{table}[htbp]
        \centering
        \begin{ruledtabular}
        \begin{tabular}{lrrrr}
         $\Delta F_a$    & \multicolumn{1}{r}{$^3P_1$}   & \multicolumn{1}{r}{$^3P_0$} & \multicolumn{1}{r}{$^1D_2$} & \multicolumn{1}{r}{$^1S_0$} \\ \hline
        Basis                   & $-73.1$            & $-80.3$      & $-67.1$      & $-569.5$                 \\
        $\Delta_{\mathrm{CBS}}$ & $1.9$              & $2.0$        & $1.6$        & 12.5                  \\
        $\Delta_{\mathrm{ref}}$ & $0.1$              & $-1.0$       & $-1.2$       & $-8.3$                    \\
        $\Delta_{\mathrm{emu}}$ & $-0.2$             & $-1.6$       & 2.2          & 4.2                    \\
        MBPT$^*$                & 0.4           & 7.0      &  15.0    &   82.6          \\
        QED$^*$                 & 1.4                   &   7.5           &   14.3           &  74.9                 \\
        Final                   & $-71.3(4)$          & $-80.9(19)$    & $-64.5(25)$    & $-561.1(93)$                
        \end{tabular}
        \end{ruledtabular}
        \caption{Calculated field shift constants (in MHz/fm$^2$) of the lowest four excited states of Ni$^{12+}$. Listed are the basis results, corrections from $\Delta_{\mathrm{CBS}}$, $\Delta_{\mathrm{ref}}$, and $\Delta_{\mathrm{emu}}$, and the MBPT and QED contributions. An asterisk (*) in the row label indicates that the corresponding effect is already included in the basis calculation, and the listed values are the difference between calculations with and without that effect. Numbers in parentheses denote the uncertainties.}
        \label{tab:final-fs}
    \end{table}
    
    Table~\ref{tab:final-fs} summarizes the final calculated differential field shift constants after including all corrections. In contrast to the behavior observed for the energy levels and the mass shift constants, QED effects play a central role in the corrections to the field shift constants. Their contributions are comparable in magnitude to those from MBPT and are generally several times larger than the corrections arising from $\Delta_{\mathrm{CBS}}$ and other CI-model-dependent terms. This strong sensitivity to QED may originates from the fact that the QED potential is concentrated near the nucleus and produces a substantial modification of the electronic wave functions in the nuclear region, precisely where the field shift constants are most sensitive.

	\begin{table*}[htbp]
        \centering
        \begin{ruledtabular}
        \begin{tabular}{lrrrrrrrr}
                                   & \multicolumn{4}{c}{$^{58}$Ni $\rightarrow$ $^{60}$Ni}                                                                                                                                                         & \multicolumn{4}{c}{$^{58}$Ni $\rightarrow$ $^{61}$Ni}                                                                                                                                                         \\ \cline{2-9} 
                                   & \multicolumn{1}{c}{$^3P_2$ $\rightarrow$ $^3P_1$} & \multicolumn{1}{c}{$^3P_2$ $\rightarrow$ $^3P_0$} & \multicolumn{1}{c}{$^3P_2$ $\rightarrow$ $^1D_2$} & \multicolumn{1}{c}{$^3P_2$ $\rightarrow$ $^1S_0$} & \multicolumn{1}{c}{$^3P_2$ $\rightarrow$ $^3P_1$} & \multicolumn{1}{c}{$^3P_2$ $\rightarrow$ $^3P_0$} & \multicolumn{1}{c}{$^3P_2$ $\rightarrow$ $^1D_2$} & \multicolumn{1}{c}{$^3P_2$ $\rightarrow$ $^1S_0$} \\ \hline
        $\Delta\nu_{\mathrm{SMS}}$ & 470(3)                                            & 268(1)                                            & 179(3)                                            & 736(10)                                           & 694(4)                                            & 396(2)                                            & 264(4)                                            & 1087(14)                                          \\
        $\Delta\nu_{\mathrm{NMS}}$ & 164(2)                                            & 176(2)                                            & 424(2)                                            & 877(3)                                            & 242(3)                                            & 260(3)                                            & 626(3)                                            & 1296(4)                                           \\
        $\Delta\nu_{\mathrm{FS}}$  & $-19$                                             & $-22(1)$                                          & $-17(1)$                                          & $-150(4)$                                         & $-25(1)$                                          & $-28(1)$                                          & $-23(1)$                                          & -196(6)                                           \\
        $\Delta\nu_{\mathrm{IS}}$  & 615(3)                                            & 422(2)                                            & 585(3)                                            & 1464(11)                                          & 912(5)                                            & 627(3)                                            & 867(5)                                            & 2187(16)                                          \\
                                   & \multicolumn{1}{l}{}                              & \multicolumn{1}{l}{}                              & \multicolumn{1}{l}{}                              & \multicolumn{1}{l}{}                              & \multicolumn{1}{l}{}                              & \multicolumn{1}{l}{}                              & \multicolumn{1}{l}{}                              & \multicolumn{1}{l}{}                              \\
                                   & \multicolumn{4}{c}{$^{58}$Ni $\rightarrow$ $^{62}$Ni}                                                                                                                                                         & \multicolumn{4}{c}{$^{58}$Ni $\rightarrow$ $^{64}$Ni}                                                                                                                                                         \\ \cline{2-9} 
                                   & \multicolumn{1}{c}{$^3P_2$ $\rightarrow$ $^3P_1$} & \multicolumn{1}{c}{$^3P_2$ $\rightarrow$ $^3P_0$} & \multicolumn{1}{c}{$^3P_2$ $\rightarrow$ $^1D_2$} & \multicolumn{1}{c}{$^3P_2$ $\rightarrow$ $^1S_0$} & \multicolumn{1}{c}{$^3P_2$ $\rightarrow$ $^3P_1$} & \multicolumn{1}{c}{$^3P_2$ $\rightarrow$ $^3P_0$} & \multicolumn{1}{c}{$^3P_2$ $\rightarrow$ $^1D_2$} & \multicolumn{1}{c}{$^3P_2$ $\rightarrow$ $^1S_0$} \\ \hline
        $\Delta\nu_{\mathrm{SMS}}$ & 911(5)                                            & 519(2)                                            & 346(5)                                            & 1425(19)                                          & 1324(8)                                           & 754(3)                                            & 504(8)                                            & 2072(27)                                          \\
        $\Delta\nu_{\mathrm{NMS}}$ & 318(4)                                            & 340(4)                                            & 821(4)                                            & 1699(6)                                           & 462(6)                                            & 495(5)                                            & 1193(5)                                           & 2470(8)                                           \\
        $\Delta\nu_{\mathrm{FS}}$  & $-34(1)$                                          & $-39(1)$                                          & $-31(1)$                                          & $-268(7)$                                         & $-43(1)$                                          & $-49(1)$                                          & $-39(2)$                                          & $-339(8)$                                         \\
        $\Delta\nu_{\mathrm{IS}}$  & 1194(7)                                           & 820(4)                                            & 1136(6)                                           & 2856(21)                                          & 1743(10)                                          & 1200(6)                                           & 1658(9)                                           & 4203(30)                                         
        \end{tabular}
        \end{ruledtabular}
        \caption{Calculated isotope shifts $\Delta\nu_{\mathrm{IS}}$ for the selected transitions in Ni$^{12+}$. The table lists the total isotope shift together with the individual contributions from the SMS term $\Delta\nu_{\mathrm{SMS}}$, the NMS term $\Delta\nu_{\mathrm{NMS}}$, and the field shift term $\Delta\nu_{\mathrm{FS}}$. Numbers in parentheses denote the estimated uncertainties.}
        \label{tab:IS}
    \end{table*}

	\begin{table}[htbp]
        \centering
        \begin{ruledtabular}
        \begin{tabular}{rrr}
        \multicolumn{1}{c}{$A$} & \multicolumn{1}{c}{$A^\prime$} & \multicolumn{1}{c}{$\delta \langle r^2 \rangle$} \\ \hline
        57.9353417              & 59.9307851                     & 0.267(5)                                         \\
                                & 60.9310548                     & 0.349(9)                                         \\
                                & 61.9283448                     & 0.478(9)                                         \\
                                & 63.9279662                     & 0.605(11)                                       
        \end{tabular}
        \end{ruledtabular}
        \caption{Atomic masses $A$ (in amu) \cite{WangAME2020Atomic2021} and nuclear radii changes $\delta \langle r^2 \rangle$ (in fm$^2$) \cite{angeliTableExperimentalNuclear2013} of several stable isotopes of Ni. Numbers in parentheses denote the uncertainties.}
        \label{tab:parameter}
    \end{table}
	
	It is important to note that the uncertainty associated with the QED corrections has not been evaluated for either the mass shift constants or the field shift constants in the present work. Recent high-precision studies of isotope shifts in HCIs, such as Ar$^{13+}$ \cite{KingOpticalAtomicClock2022}, have shown that QED effects can play a non-negligible role in mass shift constants. In particular, the QED contribution to the mass shift was found to consist of two distinct parts: a QED recoil effect and an indirect contribution arising from QED-induced shifts of energy levels, which modify the mixing of configurations. Among these, the QED recoil effect provides the dominant contribution, exceeding the latter by several times. For the field shift constants, the QED contribution was shown to arise primarily from the QED-induced modification of the mixing of configurations.

    In our calculations, QED corrections are included only through a radiative potential, which accounts for the mixing of configurations due to QED effects on the energy levels. The QED recoil effect, however, is not included. While this contribution is known to be important for an accurate description of mass shift constants, its reliable evaluation in multi-valence-electron HCIs remains a challenging problem. This limitation highlights the need for the development of improved QED potentials or effective operators capable of treating QED recoil effects in complex atomic systems.

    At the same time, the various valence-correlation effects contributing to the mass shift and field shift constants have been analyzed in detail in this work. Consequently, any residual discrepancies between our theoretical predictions and future high-precision isotope shift measurements of Ni$^{12+}$ may be attributed predominantly to missing QED recoil contributions. In this sense, the present results can serve as a benchmark for testing the accuracy and reliability of newly developed QED treatments for mass shift and field shift constants in multi-valence-electron HCIs.
	
	Based on the calculated mass shift constants and field shift constants, we obtain the isotope shifts for the lowest four excited states of Ni$^{12+}$. We evaluate the isotope shifts for the transitions from the ground state $^3\!P_2$ to the lowest four excited states, and the results are summarized in Table~\ref{tab:IS}. The nuclear parameters used in these calculations are listed in Table~\ref{tab:parameter}. As shown in Table~\ref{tab:IS}, the isotope shifts of these transitions are dominated by the mass shift, while the field shift contribution accounts for only a few percent. This is because all considered transitions lie within the fine-structure manifold of the ground-state configuration $3s^2 3p^4$ and do not involve changes of the underlying nonrelativistic orbital. The fine-structure states share similar asymptotic behavior near the nucleus, leading to a strong suppression of the field shift contribution. Consequently, these transitions are not suitable for extracting changes in the nuclear charge radius of Ni. We note that for the values in Table~\ref{tab:IS} without quoted uncertainties, the corresponding uncertainties are estimated to be below $0.5~\mathrm{MHz}$ and are negligible at the $1~\mathrm{MHz}$ level of precision after rounding.

    However, the very same suppression of the field shift makes these transitions especially attractive for searches of a possible fifth force. Isotope shift measurements aimed at probing new physics are most sensitive when the nuclear size contributions are minimized, since higher-order nuclear size effects can introduce nonlinearities in the King plot that may obscure or mimic the signatures of a fifth force.
    In the case of Yb, for example, apparent nonlinearities observed in King plot analyses were later shown to originate from higher-order nuclear size effects rather than from new physics \cite{PhysRevLett.125.123002, PhysRevLett.128.073001}.
    In contrast, the Ni$^{12+}$ fine-structure transitions considered here, particularly the two narrow and experimentally accessible visible-wavelength transitions $^3P_2$ $\rightarrow$ $^3P_1$ and $^3P_2$ $\rightarrow$ $^3P_0$, are largely insensitive to nuclear size effects and are primarily governed by the mass shift contribution. This clean separation reduces potential nuclear-structure backgrounds and makes these transitions excellent candidates for probing fifth force effects through high-precision isotope shift spectroscopy.
	
	\section{Summary and Outlook}
	
	We have performed high-precision MBPT+CI calculations of the energies and the mass shift and field shift constants of the lowest four excited states of Ni$^{12+}$. By analyzing the convergence behavior across multiple computation models and explicitly evaluating the effects of the CBS, the number of reference configurations, and the emu CI approximation, we obtained reliable excitation energies and uncertainty estimates. The final excitation energies show an average relative deviation of $0.08\%$ from experiment, with average relative uncertainties of $0.14\%$. Using the same computational framework, we determined the mass shift and field shift constants and the isotope shifts between several stable isotopes. The mass shifts are the dominant contributions to the isotope shifts of the fine-structure transitions, and the valence-correlation-induced uncertainties of the isotope shifts are below $1\%$. 

    The $^{61}$Ni isotope has a nonzero nuclear spin, and hyperfine interactions are expected to play an enhanced role in HCIs, making the evaluation of both first-order and second-order hyperfine structure important for high-precision isotope shift measurements. These contributions can be treated within the same theoretical framework as employed in this work. At the same time, the strong suppression of nuclear size effects in the isotope shifts of fine-structure transitions of Ni$^{12+}$ makes these transitions particularly well suited for King plot analyses aimed at probing a possible fifth force, as they avoid nonlinearities arising from higher-order nuclear size effects. Ni$^{12+}$ therefore emerges as a promising platform for precision measurements and tests of fundamental physics.

	\section{acknowledgment}

	 We acknowledge support from the Strategic Priority Research Program of the Chinese Academy of Sciences (Grant No. XDB0920403) and the National Key Research and Development Program of China (Grant No. 2022YFB3904002).


	
	\bibliography{ref}
	
\end{document}